\documentclass[fleqn,a4paper,12pt]{article}
\usepackage{amssymb}
\usepackage{makeidx}
\usepackage{amsmath}
\usepackage{graphicx}
\usepackage{lscape}
\usepackage{epsfig}

 \newcommand{\newc}{\newcommand}
\newc{\beq}{\begin{equation}} \newc{\eeq}{\end{equation}}
\newc{\bea}{\begin{array}} \newc{\eea}{\end{array}}
\newc{\ri}{{\mathrm i}}
\newc{\bW}{{\mathbf W}}
\newc{\bR}{{\mathbf R}}
\newc{\bN}{{\mathbf N}}
\newc{\Psibar}{\overline\Psi}
\newc{\w}{{\bf w}}
\newc{\E}{{\mathbf{E}}}
\newc{\bp}{{\bf p}}
\newc{\ta}{\tilde a}
\newc{\bV}{{\bf V}}
\newc{\bfV}{{\bf V}}
\newc{\bfG}{{\bf G}}
\newc{\bx}{{\bf x}}
\newc{\bu}{{\bf u}}
\newc{\bP}{{\bf P}}
\newc{\bJ}{{\bf J}}
\newc{\bK}{{\bf K}}
\newc{\pd}{{\partial}}
\newc{\ti}{{\times}}
\newc{\bA}{{\bf A}}
\newc{\bE}{{\bf E}}
\newc{\bfn}{{\bf\nabla}}
\newc{\ho}{\hookrightarrow}
\newc{\ra}{\rightarrow}
\newc{\bv}{{\bf v}}
\newc{\bb}{{\bf b}}
\newc{\bc}{{\bf c}}
\newc{\bd}{{\bf d}}
\newc{\tbb}{\tilde{\bf b}}
\newc{\tbc}{\tilde{\bf c}}
\newc{\tbd}{\tilde{\bf d}}
\newc{\bz}{{\bf 0}}
\newc{\bun}{{\bf 1}}
\newc{\bL}{{\bf L}}
\newc{\bS}{{\bf S}}
\newc{\bB}{{\bf B}}
\newc{\br}{{\bf r}}
\newc{\sig}{{\mathbf\sigma}}
\newc{\eg}{{\it e.g.\ }}
\newc{\bpi}{{\mathbf\pi}}
\newc{\ie}{{\it i.e.\ }}
\newc{\etal}{{\it et al}}

\def\JPA#1#2#3#4{#2 #1 {\em J. Phys. A: Math. Gen.} {\bf #3} #4}

\def\AP#1#2#3#4{#2 #1 {\em Ann. Phys.} {\bf #3} #4}

\def\NCB#1#2#3#4{#2 #1 {\em Nuov. Cim.} {\bf #3 B} #4}

\def\IJTP#1#2#3#4{#2 #1 {\em Int. J. Theor. Phys.} {\bf #3} #4}
\def\CMP#1#2#3#4{#2 #1 {\em Comm. Math. Phys.} {\bf #3} #4}

\def\PNASUS#1#2#3#4{#2 #1 {\em Proc. Nat. Acad. Sci. U.S.} {\bf #3} #4}
\def\TMP#1#2#3#4{#2 #1 {\em Theor. Math. Phys.} {\bf #3} #4}

\def\TMP#1#2#3#4{#2 #1 {\em Theor. Math. Phys.} {\bf #3} #4}

\catcode`@=11 \long
\def\@caption#1[#2]#3{\par\addcontentsline{\csname
ext@#1\endcsname}{#1} {\protect\numberline{\csname
the#1\endcsname}{\ignorespaces #2}} \begingroup \small
\@parboxrestore \@makecaption{\csname fnum@#1\endcsname}
{\ignorespaces #3}\par \endgroup} \catcode`@=12

\begin{document}
 \begin{titlepage} \vskip 2cm
\begin{center} {\Large\bf \Large\bf Galilei invariant theories.\\III.
Wave equations for massless fields} \footnote{E-mail: {\tt
niederle@fzu.cz},\ \ {\tt nikitin@imath.kiev.ua} } \vskip 3cm {\bf
J. Niederle$^a$, \\ and A.G. Nikitin$^b$ } \vskip 5pt {\sl
$^a$Institute of Physics of the Academy of Sciences of the Czech
Republic,\\ Na Slovance 2, 18221 Prague, Czech Republic} \vskip 2pt
{\sl $^b$Institute of Mathematics, National Academy of Sciences of
Ukraine,\\ 3 Tereshchenkivs'ka Street, Kyiv-4, Ukraine, 01601\\}
\end{center}
\vskip .5cm \rm

\begin{abstract}Galilei invariant equations for massless fields are obtained via
contractions of relativistic wave equations. It is shown that the collection
 of non-equivalent Galilei-invariant wave equations for massless fields with
 spin equal 1 and 0 is very broad and describes
 many physically consistent systems.
 In particular, there exist a huge number of such equations for massless
 fields which correspond to various contractions of representations of the
 Lorentz group to those of the Galilei one.
\end{abstract}

\end{titlepage}

\setcounter{footnote}{0} \setcounter{page}{1}
\setcounter{section}{0}

\section{Introduction}

It was observed by Le Bellac and L\'evy-Leblond \cite{lebellac}
already in 1973
 that the non-relativistic limit of the Maxwell
equations is not defined in a unique way. According to Le Bellac and
L\'evy-Leblond (\cite{lebellac} p. 218), the term "non-relativistic"
means "in agreement with the principle of Galilean relativity".
Moreover, they claimed that there exist two Galilei invariant
theories of electromagnetism which can be obtained by appropriate
limiting procedures starting with the Maxwell theory.

The combination of word "Galilean electromagnetism" itself
introduced in \cite{lebellac} looked rather strange since it is
pretty well known that electromagnetic phenomena are in perfect
accordance with the Einstein relativity principle. However
physicists are always interested when non-relativistic
approximations are adequate, which makes the results of paper
\cite{lebellac} quite popular. The importance of such result is
emphasized by the fact that the correct definition of
non-relativistic limit is by no means a simple problem in general
and in the case of theories of massless fields in particular, see,
for example, \cite{Hol}.

Analyzing the contents of main impact journals in theoretical and
mathematical physics one finds  that the interest of researches in
Galilean aspects of electrodynamics belongs to evergreen subjects.
Various approaches to Galilei invariant theories which were
discussed briefly in \cite{NN1}. An effective approach to Galilean
electromagnetism was used in papers \cite{ijtp}, \cite{santos} in
which Maxwell equations were generalized in to (1+4)-dimensional
Minkowski space and then reduced to Galilei invariant equations.
Such reduction is based on the fact that the Galilei group is a
subgroup of the generalized Poincar\'e group (i.e., of the group of
motions of the flat (1+4)-dimensional Minkowski space). For
reduction of representations of the group $P(1,4)$ to those of the
Galilei group see refs. \cite{FN1} and \cite{FN2}.

Nevertheless, the Galilean electromagnetism still contains many
unsolved problems, for example, the question of complete description
of all possible Galilean theories for vector and scalar massless
fields. And exactly solution of this problem is the main issue of
the present paper.

In paper \cite{NN1} indecomposable representations of homogeneous
Galilei group  $HG(1,3)$ were derived, namely all those which when
restricted to representations of the rotation subgroup of the group
$HG(1,3)$, are decomposed to spin 0 and spin 1 representations.
Moreover, their connection with representations of the Lorentz group
via the In\"on\"u-Wigner contractions \cite{contraction} were
studied in \cite{NN1} and \cite{NN2}. These results open a way to
complete classification of the wave equations describing the related
fields both massive and massless ones.

In the present paper we use our knowledge of indecomposable
representations of the homogeneous Galilei algebra $hg(1,3)$ to
figure out the Galilei invariant equations for vector and scalar
massless fields. We shall show that in contrast to the corresponding
relativistic equations for which there are only two possibilities --
the Maxwell equations and equations for the longitudal massless
field, the number of possible Galilean equations is very huge. Among
them there are equations with more component and less component
fields then in the Maxwell equation.

These results can be clearly interpreted in terms of representation
and contraction theories. As it was proved in papers \cite{NN1} and
\cite{NN2}, there is a large variety of possible contractions of
representations of the Lorentz group to those of Galilei one, and,
consequently many non-equivalent Galilean massless fields. In the
following sections we use these results of \cite{NN1}, \cite{NN2} to
describe connections of relativistic and Galilean formulations of
theories for massless fields. These connections appear to be rather
non-trivial: in particular, completely decoupled relativistic
systems could be contracted to coupled Galilean ones.

In Sections 2 and 3 we present some results of paper \cite{NN1}
related to the classification of indecomposable representations of
the homogeneous Galilei group and contractions of the related
representations of Lorentz group. These results are used in Sections
4 and 5 to classify all non-equivalent Galilei-invariant equations
for massless fields. Then the classification results are summarized
and discussed in Section 6.


\section{Indecomposable representations of the homogeneous Galilei group}

The Galilei group $G(1,3)$ consists of the following transformations
of temporal and spatial variables: \beq\label{11}\bea{l} t\to
t'=t+a,\\{\bf x}\to {\bf x}'={\bf R}{\bf x} +{\bf v}t+\bf b,\eea\eeq
where $a,{\bf b}$ and ${\bf v}$ are real parameters of time
translation, space translations and pure Galilei transformations
respectively, and $\bf R$ is a rotation matrix.

The homogeneous Galilei group $HG(1,3)$ is a subgroup of $G(1,3)$
leaving invariant a point $\textbf{x}=(0,0,0)$ at time $t=0$. It is
formed by space rotations and pure Galilei transformations, i.e., by
transformations (\ref{11}) with $a={\bf b}\equiv 0$.

The Lie algebra $hg(1,3)$ of the homogeneous Galilei group includes
six basis elements, i.e., three generators $S_a, a=1,2,3$ of the
rotation subgroup and three generators $\eta_a$ of Galilean boosts.
These basis elements should satisfy the following commutation
relations
 \beq\label{e3}\bea{l}
[S_a,S_b]=\ri\varepsilon_{abc}S_c,\\
    {[}\eta_a,S_b{]}=\ri\varepsilon_{abc}\eta_c,\\
{[}\eta_a,\eta_b{]}=0.\eea\eeq

 All indecomposable representations of $HG(1,3)$
which when restricted to the subgroup of rotations are decomposed to
direct sums of vector and scalar representations, has been found in
paper \cite{NN1}. These indecomposable representations are labeled
by triplets of numbers $n,m,\lambda$ and denoted as
$D(m,n,\lambda)$). The labeling numbers take the values
\beq\label{mn}\bea{l}-1\leq
(n-m)\leq2, \ n\leq3, \\\\ \lambda=\left\{\bea{l}0\ \texttt{if}\ m=0,\\
1\ \texttt{if}\ m=2\ \texttt{or}\ n-m=2,\\ 0, 1\ \texttt{if}\ m=1,
n\neq3. \eea\right.\eea\eeq

In accordance with (\ref{mn}) there exist ten non-equivalent
indecomposable representations $D(m,n,\lambda)$). The spaces of
 representation  $D(m,n,\lambda)$)
 can include
 three types of rotational scalars $A, B, C$  and five
 types of vectors ${\bf R}, {\bf U}, {\bf W}, {\bf K}, {\bf N}$ whose
 transformation laws with respect to
 the Galilei boost are:
 \beq\label{fin}\bea{l}
 A\to A'=A,\\B\to B'=B+{\bf v \cdot R},\\C\to C'=C+{\bf v\cdot U}+\frac12{\bf v}^2 A,\\
 {\bf R}\to{\bf R}'={\bf R},\\{\bf U}\to{\bf U}'={\bf U}+{\bf v}A,
 \\{\bf W}\to {\bf W}'={\bf W}+{\bf v}\times{\bf R},\\
 {\bf K}\to{\bf K}'={\bf K}+{\bf v}\times{\bf R}+{\bf v}A,\\
 {\bf N}\to{\bf N}'={\bf N}+{\bf v}\times{\bf W}+{\bf v}B+
 {\bf v}({\bf v\cdot\bf R})-\frac12{\bf v}^2{\bf R}\eea\eeq
 where ${\bf v}$ is a vector whose components are parameters of the Galilei
 boost, ${\bf v \cdot R}$ and ${\bf v}\times{\bf R}$ are
 scalar and vector products of vectors $\bf v$ and $\bf R$.

The carrier spaces of indecomposable representations of group
$hg(1,3)$ include such sets of scalars $A, B, C$
 and vectors ${\bf R}, {\bf U}, {\bf W}, {\bf K}, {\bf N}$
 which transform between themselves w.r.t. the transformations
 (\ref{fin}) but cannon be split to a direct sum of invariant subspaces.
  There exist exactly ten such sets:
  \beq\label{rep}\bea{rcl}\{A\}& \Longleftrightarrow&\ D(0,1,0),\\
  \{{\bf R}\}& \Longleftrightarrow&\ D(1,0,0),\\\{B,  {\bf R}  \}&\
  \Longleftrightarrow&\ D(1,1,0),\\\{A, {\bf U}\}&\ \Longleftrightarrow&\
D(1,1,1),\\
  \{A,{\bf U}, C\}&\ \Longleftrightarrow&\ D(1,2,1),\\
\{{\bf W},
 {\bf R}\}&\ \Longleftrightarrow&\ D(2,0,0), \\
 \{{\bf R, W}, B\}& \ \Longleftrightarrow&\ D(2,1,0),\\
  \{A, {\bf K}, {\bf R}\}&\ \Longleftrightarrow&\ D(2,1,1),\\
   \{A, B, {\bf K}, {\bf R}\}&\ \Longleftrightarrow&\ D(2,2,1),\\
   \{A, {\bf N},{\bf W}, {\bf R}\}&\ \Longleftrightarrow&\ D(3,1,1).\eea
 \eeq

Thus in contrary to the relativistic case where are only three
Lorentz covariant quantities which transforms as vectors or scalars
under rotations, i.e., the relativistic four-vector, antisymmetric
tensor of second order and scalar, there are ten indecomposable sets
of Galilean vectors and scalars which we enumerate in equation
(\ref{rep}).

\section{Contractions of representations of the Lorentz algebra}

It is well known that the Galilei algebra can be obtained from the
Poincar\'e algebra by a limiting procedure called "the
In\"on\"u-Wigner contraction" \cite{contraction}. Representations of
these algebras also can be connected by this kind of contraction.
However, this connection is more complicated in two reasons. First,
the contraction of a non-trivial representation of the Lorentz
algebra yields to the representation of the homogeneous Galilei
algebra which the generators of Galilei boost are represented
trivially, so to obtain a non-trivial representation it is necessary
to apply additionally a similarity transformation which depends on
the contraction parameter in a tricky way. Second, to obtain an
indecomposable representations of $hg(1,3)$, it is necessary in
general to start with completely reducible representations of the
Lie algebra of Lorentz group.

In paper \cite{NN1} representations of the Lorentz group which can
be contracted to representations
 $D(m,n,\lambda)$ of the Galilei group were found
 and the related contractions where specified. Here we present only
 a part of the results from \cite{NN1} which will be used in what follows.

Let us start with the representation $D(\frac12,\frac12)$ of the Lie
algebra $so(1,3)$ of the Lorentz group, whose carrier space is
formed by four-vectors. Basis of this representation is given by
$4\times4$ matrices of the following form:
\beq\label{repa}S_{ab}=\varepsilon_{abc}\left(\bea{cc}s_c&\bz_{3\times1}\\
\bz_{1\times3}&0\eea\right),\ \ S_{0a}=
\left(\bea{cc}\bz_{3\times3}&- k^\dag_a\\k_a&0\eea\right).\eeq Here
$s_a$ are
 matrices of spin one with elements $(s_a)_{bc}=i
\varepsilon_{abc}$, $k_a$ are $1\times3$ matrices of the form
\beq\label{k} k_1=\left (\ri, 0, 0\right),\qquad k_2=\left (0, \ri,
0\right), \qquad k_3=\left (0, 0, \ri\right).\eeq

The In\"on\"u-Wigner contraction consists of transformation to a new
basis
\[S_{ab}\to S_{ab}, \ S_{0a}\to\varepsilon S_{0a}\]
followed by a similarity transformation of all basis elements
$S_{\mu\nu}\to S'_{\mu\nu}=VS_{\mu\nu}V^{-1}$ with a matrix $V$
depending on contracting parameter $\varepsilon$. Moreover, $V$
should depend on $\varepsilon$ in a tricky way, such that all the
transformed generators $S'_{ab}$ and $\varepsilon S'_{0a}$ are kept
non-trivial and non-singular when $\varepsilon\to 0$
\cite{contraction}.

There exist two matrices $V$ for representation (\ref{repa}), namely
\beq\label{con1}V_1=\left(\bea{rc}\varepsilon
I_{3\times3}&\bz_{3\times1}\\
\bz_{1\times3}&1\eea\right),\eeq and
\beq\label{con2}V_2=\left(\bea{rc}
I_{3\times3}&\bz_{3\times1}\\
\bz_{1\times3}&\varepsilon\eea\right).\eeq

Using (\ref{con1}) we obtain
\beq\label{con3}S'_{ab}=V_1S_{ab}V_1^{-1}=S_{ab},\
S'_{0a}=\varepsilon
V_1S_{0a}V_1^{-1}=\left(\bea{cc}\bz_{3\times3}&-\varepsilon^2k^\dag_a
\\k_a&0\eea\right).\eeq
Then, passing $\varepsilon$ to zero, we come to the following
matrices \beq\label{con4}S_a=\frac12\varepsilon_{abc}S_{bc}=
\left(\bea{cc}s_a&\bz_{3\times1}\\
\bz_{1\times3}&0\eea\right),\ \eta_a=\lim
S'_{0a}|_{\varepsilon\to0}=
\left(\bea{cc}\bz_{3\times3}&\bz_{3\times1}
\\k_a&0\eea\right).\eeq

Analogously, using matrix $V_2$ instead of $V_1$ we obtain
\beq\label{con5}S_a=
\left(\bea{cc}s_a&\bz_{3\times1}\\
\bz_{1\times3}&0\eea\right),\ \eta_a=
\left(\bea{cc}\bz_{3\times3}&-k^\dag_a
\\\bz_{1\times3}&0\eea\right).\eeq

Matrices (\ref{con4}) and (\ref{con5}) satisfy commutation relations
(\ref{e3}), i.e., they realize representations of algebra $hg(1,3)$.
They are generators of indecopmosable representations $D(1,1,0)$ and
$D(1,1,1)$ of the homogeneous Galilei group respectively. Indeed,
denoting vectors from the related representation spaces as
\[\Psi=\left(\bea{c}R_1\\R_2\\R_3\\B\eea\right) \ \ \texttt{for }
D(1,1,0)\ \ \texttt{and
}\tilde\Psi=\left(\bea{c}U_1\\U_2\\U_3\\A\eea\right) \ \ \texttt{for
} D(1,1,1)\] and using the transformation laws (\ref{fin}) for $A,\
B,\  {\bf R}=\texttt{column}(R_1,R_2,R_3) $ and $ \ {\bf
U}=\texttt{column}(U_1,U_2,U_3)$ we easily find the corresponding
Galilei boost generators $\eta_a$ in the forms (\ref{con4}) and
(\ref{con5}). As far as rotation generators $S_a$ are concerned
they are direct sums of matrices of spin one (which are responsible
for transformations of 3-vectors $\bf R$ and $\bf U$) and zero
matrices (which keep scalars $A$ and $B$ invariant).

To obtain five-dimension representation $D(1,2,1)$ we are supposed
to start with a direct sum of representations $D(\frac12,\frac12)$
and $ D(0,0)$ of the Lorentz group. The corresponding generators of
algebra $so(1,3)$ have the form \beq\label{Smu}\hat
S_{\mu\nu}=\left(\bea{cc}S_{\mu\nu}&\cdot\\\cdot&0\eea\right),\eeq
where $S_{\mu\nu}$ are matrices (\ref{repa}) and the dots denote
zero matrices of an appropriate dimension. The corresponding
 matrix of similarity transformation can be written as:
\beq\label{mat}V_3=\left(\bea{llr}
I_{3\times3}&0_{3\times1}&0_{3\times1}\\0_{1\times3}&\frac12\varepsilon&\frac12\varepsilon
\\0_{1\times3}&-\varepsilon^{-1} &\varepsilon^{-1}\eea\right),\ \
U_3^{-1}=\left(\bea{llr}
I_{3\times3}&0_{3\times1}&0_{3\times1}\\0_{1\times3}&\varepsilon^{-1}&-\frac12\varepsilon
\\0_{1\times3}&\varepsilon^{-1} &\frac12\varepsilon\eea\right).\eeq

As a result we obtain the following basis elements of representation
$D(1,2,1)$ of algebra $hg(1,3)$:
\beq\label{con7}S_a=\left(\bea{ccc}s_a&\bz_{1\times3}
&\bz_{1\times3}\\\bz_{3\times1}&0&0\\
    \bz_{3\times1}&0&0\eea\right),\ \ \
    \eta_a=\left(\bea{ccc}\bz_{3\times3}&k^\dag_a&\bz_{3\times1}\\
  \bz_{1\times3}&0&0\\
    k_a&0&0\eea\right).\eeq

    Matrices $\eta_a$ (\ref{con7}) generate transformations
    (\ref{fin}) of components of vector-function
    $\hat\Psi=\texttt{column}({\bf U}, A, C)$ so that the relation
    \[\hat\Psi\to\hat\Psi'=
    \exp(\ri{\mbox{\boldmath$\eta\cdot v$\unboldmath}})\hat\Psi\]
    being written componentwise, simply coincides with the corresponding
    transformation properties written in
    (\ref{fin}).

In paper \cite{NN1} was shown  how all the other representations
$D(m,n,\lambda)$ of homogeneous Galilei algebra can be obtained via
special contractions of appropriate representations of Lorentz
algebra. Here we have presented only such contractions which will be
used later on.

\section{Galilean massless fields}

To construct Galilean massless equations it is possible to use the
same approach which we have used in \cite{NN3} to derive equations
for the massive fields. However we prefer to apply another technique
which consists in  contractions of the relativistic wave equations.

\subsection{Galilean limits of Maxwell's equations}

According to the analysis made by  L\'evy-Leblond and Le Bellac in
1967 \cite{ll1967}, \cite{lebellac} (see also
\cite{levyleblond},\cite{levycarroll}) there are two Galilean limits
of Maxwell's equations.

In the so called "magnetic" Galilean limit we recieve a
pre-Maxwellian electromagnetism. The corresponding equations for a
magnetic field $\bf H$ and electric field $\bf E$ take the form
\beq\label{mag}\bea{l}\nabla\times {\bf E}_m-\frac{\partial {\bf
H}_m}{\partial t}=0, \ \ \nabla\cdot{\bf
E}_m=ej^0_m,\\
\nabla\times {\bf H}_m=e{\bf j}_m,\ \ \ \ \ \ \ \ \nabla\cdot{\bf
H}_m=0\eea\eeq where $j=(j^0_m,{\bf j}_m)$ is an electric current
and $e$ is electric charge.

Equations (\ref{mag}) are invariant with respect to the Galilei
transformations (\ref{11}) provided vectors ${\bf H}_m, \ {\bf E}_m$
and electric current $j$ cotransform as \beq\label{tran}\bea{l}{\bf
H}_m\to{\bf H}_m,\ {\bf E}_m \to {\bf E}_m-{\bf v}\times{\bf
H}_m,\\{\bf j}_m\to{\bf j}_m,\ \ \ \ j^0_m\to j^0_m+{\bf v}\cdot{\bf
j}_m.\eea\eeq

Introducing Galilean vector-potential $A=(A^0,{\bf A})$ such as
\beq\label{pot}{\bf H}_m=\nabla\times{\bf A},\ \ {\bf
E}_m=-\frac{\partial {\bf A}}{\partial t}-\nabla A^0\eeq we conclude
that the related transformation lows for $A$ have the following
form: \beq A^0\to A^0+{\bf v}\cdot{\bf A}, \ {\bf A}\to{\bf
A}.\label{A^0}\eeq

The other, "electric" Galilean limit of Maxwell's equations looks as
\beq\label{el}\bea{l}\nabla\times {\bf H}_e+\frac{\partial {\bf
E}_e}{\partial
t}=e{\bf j}_e, \ \ \nabla\cdot{\bf E}_e=ej^4_e,\\
\nabla\times {\bf E}_e=0,\ \ \ \ \ \ \ \ \ \ \ \ \ \nabla\cdot{\bf
H}_e=0,\eea\eeq whilst the Galilean transformation lows have the
following form \beq\label{t}\bea{l}{\bf H}_e\to{\bf H}_e+{\bf
v}\times{\bf E}_e,\ \ {\bf E}_e\to{\bf E}_e,
\\{\bf j}_e\to{\bf j}_e+{\bf v}j^4_e,\ \ \ \ \ \ \ \ \ \ \
j^4_e\to j^4_e.\eea\eeq Vectors ${\bf H}_e$ and ${\bf E}_e$ can be
expressed via vector-potentials as follows \beq\label{po}{\bf
H}_e=\nabla\times{\bf A},\ \ \ {\bf E}_e=-\nabla A^4\eeq and so the
corresponding Galilei transformations for the vector-potential are:
\beq A^4\to A^4, \ {\bf A}\to{\bf A}+{\bf v}A^4;\label{A^4}\eeq

Till this point we just presented the results of the L\'evy-Leblond
analysis of two possible Galilean limits of Maxwell's equations. Let
us note that these results admit clear interpretation using the
representation theory. Indeed, in accordance with the results
presented in Section 3 there exist exactly two non-equivalent
representations of the homogeneous Galilei group the carrier spaces
of which are four-vectors -- the representations $D_1(1,1,0)$ and
$D_1(1,1,1)$. In other words there are exactly two non-equivalent
Galilei transformations for four-vector-potentials and currents,
which are given explicitly by equations (\ref{A^0}), (\ref{tran})
and (\ref{A^4}), (\ref{t}). Equations for massless fields invariant
with respect to these transformations are given by relations
(\ref{mag}) and (\ref{el}) respectively.

Both representations, i.e.,  $D_1(1,1,0)$ and $D_1(1,1,1)$, can be
obtained via contractions of the representation $D(1/2,1/2)$ of the
Lorentz group whose carrier space is formed by relativistic
four-vectors. The related contraction matrices are given explicitly
by relations (\ref{con1}) and (\ref{con2}). Each of these
contractions generates the Galilean limit of Maxwell's equations,
and in this way we obtain easily the systems (\ref{mag}) and
(\ref{el}). In the following section we will obtain equations
(\ref{mag}) and (\ref{el}) via contraction of a more general system
of relativistic equations for massless fields.

\subsection{Extended Galilean electromagnetism}

In accordance with our analysis of vector field representations of
the Galilei group there exists the only representation, namely,
$D_1(1,2,1)$ whose carrier space is formed by five-vectors. Such
five-vectors appears naturally in many Galilean models, especially
in those ones which are constructed via {\it reduction approach}
\cite{NN1}, i.e., starting with models invariant with respect to the
extended Poincar\'e group $P(1,4)$  and making reduction to its
Galilean subgroup.

As mentioned in \cite{NN1}, there is a formal possibility to
introduce such five component gauge fields which join and extend the
magnetic and electric Galilean limits of relativistic four--vector
potentials. However the physical meanings of the related theories
was not clear. Moreover, Maxwell's electrodynamics can be contracted
either to the magnetic (\ref{mag}) or to the electric limit
(\ref{el}), and it is generally accepted to think that it is
impossible to formulate a consistent theory including both
`electric' and `magnetic' types of Galilean gauge fields, see, e.g.,
\cite{santos}.

In contrary to \cite{santos}, we shall show that it is possible to
join the `electric' and `magnetic' Galilean gauge fields since the
Galilean five-vector potential appears naturally via contraction of
a relativistic theory. Rather surprisingly, the related relativistic
equations can be decoupled to two non-interacting subsystems whereas
its contracted counterpart appears to be coupled. This is in
accordance with the observation presented in \cite{NN1} that some
indecomposable representations of the homogeneous Galilei group
appears via contractions of the completely reducible representations
of the Lorentz group.

Let us start with relativistic equations for the vector-potential
$A^\mu$  \beq\label{relpot}p^\mu p_\mu A^\nu=-ej^\nu\eeq in the
Lorentz gauge \beq\label{Lorentz}p_\mu A^\mu=0 \ \ \texttt{or
}p_0A^0={\bf p}\cdot{\bf A}.\eeq

Consider also the inhomogeneous d'Alembert equation for a
relativistic scalar field which we denote as $A^4$:
\beq\label{scalpot}p^\mu p_\mu A^4=ej^4.\eeq

Introducing the related vectors of the field strengthes in the
standard form \beq\label{F}  {\bf H}=\nabla\times{\bf A},\ \ {\bf
E}=-\frac{\partial {\bf A}}{\partial x_0}-\nabla A^0, \ {\bf
F}=\nabla A^4, \ F^0=\frac{\partial A^4}{\partial x_0}\eeq we come
to Maxwell's equations for ${\bf E}$ and ${\bf H}$: \beq\label{HH}
\bea{l}\displaystyle\nabla\times {\bf E}-\frac{\partial {\bf
H}}{\partial x_0}=0, \ \ \nabla\cdot{\bf H}=0,\\\\\displaystyle
\nabla\times {\bf H}+\frac{\partial {\bf E}}{\partial x_0}=e{\bf
j},\ \ \ \ \ \ \nabla\cdot{\bf E}=ej^0\eea\eeq and the following
equations for $\bf F$ and $F^0$
\beq\label{FF}\bea{l}\displaystyle\frac{\partial { F^0}}{\partial
x_0}+\nabla\cdot{\bf F}=ej^4,\\\\\displaystyle \nabla\times{\bf
F}=0,\ \ \frac{\partial {\bf F}}{\partial x_0}=\nabla F^0.\eea\eeq

Surely the system of equations (\ref{HH}) and ({\ref{FF}) is
completely decoupled. Rather surprisingly its Galilean counterpart
which will be obtained using the In\"on\"u-Wigner contraction
appears to be coupled. This contraction can be made directly for
equations (\ref{HH}), (\ref{FF}) but we prefer a more simple way
with using the potential equations (\ref{relpot}).

The system of equations (\ref{relpot})-(\ref{scalpot}) is a
decoupled system of relativistic equations for the five component
function \beq\label{A}{\hat A}=\texttt{column}(A^1,A^2,A^3,A^0,A^4)=
\texttt{column}({\bf A},A^0,A^4).\eeq Moreover, the components
$(A^1,A^2,A^3,A^0)$ transform as four-vector and $A^4$ transforms as
a scalar, so the generators $ S_{\mu \nu}$ of the Lorentz group
defined on $A$ are direct sums of the four-vector generators $\hat
S_{\mu\nu}$ and zero matrices given by equation (\ref{Smu}). In
other words, such generators realize the direct sum
$D(\frac12,\frac12)\oplus D(0,0)$ of representations of algebra
$so(1,3)$.

In accordance with results of paper \cite{NN1} the completely
reduced representation of the Lie algebra of Lorentz group whose
basis elements are given by equation (\ref{Smu}) can be contracted
either to direct sum of indecomposable representations of  the the
Galilei algebra $hg(1,3)$ or to indecomposable representation
$D_1(1,2,1)$ this algebra.

Let us consider the first possibility, i.e., the contraction to the
indecomposable representation. Such contraction is presented in
Section 3, see equations (\ref{Smu})--(\ref{con7}) here.

Let us demonstrate that this contraction reduces the decoupled
relativistic system (\ref{relpot}), (\ref{scalpot}) to a system of
coupled equations invariant with respect to the Galilei group.
Indeed denoting $A'=U_3A=({\bf A}',A'^0,A'^4)$ and $j'=U_3j= ({\bf
j}'j'^0,j'^4)$ and taking into account that the relativistic
variable $x_0$ is related to non-relativistic variable $t$ as
$x_0=ct$, we come to the following system of equations for the
transformed quantities: \beq\label{sys1}{\bf p}^2  A'^k=-e j'^k, \ \
\  \ri \frac{\partial A'^4}{\partial t}={\bf p}\cdot{\bf A}'.\eeq

Generators of Galilei group for vectors $A'$ and $j'$ are given by
equation (\ref{con7}), so under the Galilei transformation
(\ref{11}) they cotransform in accordance with the representation
$D_1(1,2,1)$, i.e., \beq\label{trans}A^0\to A^0+{\bf v}\cdot {\bf
A}+\frac{{\bf v}^2}2A^4,\ {\bf A}\to{\bf A}+{\bf v}A^4,\ A^4\to
A^4,\eeq and
 \beq \label{16}j^4\to j^4, \ {\bf j}\to{\bf
j}+{\bf v}j^4,\ j^0\to j^0+{\bf v}\cdot{\bf j}+\frac12v^2j^4.\eeq

Of course transformations (\ref{11}), (\ref{trans}) and (\ref{16})
keep the system (\ref{sys1}) invariant. In accordance with
(\ref{sys1}) the related field strengthes (compare with (\ref{F}))
\beq\label{RWNB}{\bf W}=\nabla\times{\bf A}',\ \ {\bf
N}=-\frac{\partial {\bf A}'}{\partial t}-\nabla  A'^0,\ \ {\bf
R}=\nabla  A'^4, \ \ B=\frac{\partial  A'^4}{\partial t}\eeq satisfy
the following equations \beq\label{coupl}\bea{l}{\cal
J}^0\equiv\nabla\cdot{\bf N}-\frac{\partial}{\partial
t}B-ej^0=0,\\{\mbox{\boldmath${\cal J}$\unboldmath}} \equiv
\nabla\times{\bf W}+\nabla B-e{\bf j}=0,\\{\cal
J}^4\equiv\nabla\cdot{\bf R}-ej^4=0,\\{\mbox{\boldmath${\cal
N}$\unboldmath}}\equiv\frac{\partial}{\partial t}{\bf
W}+\nabla\times{\bf N}=0,\\{\mbox{\boldmath${\cal
W}$\unboldmath}}\equiv\frac{\partial}{\partial t}{\bf R}-\nabla
B=0,\\ {\mbox{\boldmath${\cal R}$\unboldmath}}\equiv \nabla\times
{\bf R}=0,\\ {\cal B}\equiv\nabla\cdot {\bf W}=0.\eea\eeq

Like (\ref{sys1}), equations (\ref{coupl}) are covariant with
respect to the Galilei group. Moreover, the Galilei transformations
for fields ${\bf R}, {\bf W}, {\bf N}$ , $B$ and current $j$ are
given by equations (\ref{fin}) and (\ref{16}) respectively.
 In other words, these fields and the current $j$ form carrier spaces
of representation $D_1(3,1,1)$ and $D_1(1,2,1)$ of algebra $hg(1,3)$
correspondingly.

In contrast with the decoupled relativistic system of equations
(\ref{HH}) and (\ref{FF}) its Galilean counterpart (\ref{coupl})
appears to be a coupled system of equations for vectors ${\bf R},
{\bf W},\ {\bf N}$ and scalar $B$.

The system of equations (\ref{coupl}) was obtained in paper
\cite{ijtp} by reduction of generalized Maxwell equations invariant
with respect to extended Poincar\'e group $P(1,4)$ including one
time variable and four spatial variables. We prove that this system
is nothing but the contracted version of system (\ref{HH}),
(\ref{FF}) including the ordinary Maxwell equations and equations
for four-gradient of scalar potential. It other words, the system of
Galilei invariant equations (\ref{coupl}) admits a clear physical
interpretation as a non-relativistic limit of the system of familiar
equations (\ref{HH}) and (\ref{FF}).

\subsection{Reduced Galilean electromagnetism}

In contrast with the relativistic case the Galilei invariant
approach makes it possible to reduce the number of field variables.
For example, considering the magnetic limit (\ref{mag}) of the
Maxwell equations it is possible to restrict ourselves to the case
${\bf H}_m=0$ in as much as this condition is invariant with respect
to the Galilei transformations (\ref{tran}). Notice that in the
relativistic theory such condition can be imposed only in a
particular frame of references and will be affected by the Lorentz
transformation.

In the mentioned sense equations (\ref{coupl}) are reducible too.
They are defined on the most extended multiplet of vector and spinor
fields which is a carrier space of indecomposable representation of
the homogeneous Galilei group. The related representation
$D_1(3,1,1)$ is indecomposable but reducible, i.e., includes
subspaces invariant with respect to the Galiei group. This makes it
possible to reduce the number of dependent components of equations
(\ref{coupl}) without violating its Galilei invariance.

In this section we consider systematically all possible Galilei
invariant constrains which can be imposed on solutions of equations
(\ref{coupl}) and present the related reduced versions of Galilean
electromagnetism.

In accordance with (\ref{fin}) and (\ref{16}) the vector $\bf R$ and
the fourth component $j^4$ of current form  invariant subspaces with
respect to Galilei transformations. Thus we can impose the
Galilei-invariant conditions \beq\label{V=0}{\bf R}=0\ \ \texttt{or
}\nabla A^4=0,  \ \ j^4=0\eeq and reduce the system (\ref{coupl}) to
the following one \beq\label{coupl1}\bea{l}\frac{\partial}{\partial
t}\tilde{\bf H}+\nabla\times\tilde{\bf E}=0,\\\nabla\times\tilde{\bf
H}=e{\bf j},\ \nabla\cdot \tilde{\bf H}=0,\\\nabla\cdot\tilde{\bf
E}=\frac{\partial}{\partial t}S+ej^0,\\\nabla S=0\eea\eeq where we
denote $\tilde{\bf H}={\bf W}|_{{\bf R}\equiv0},\ \tilde{\bf E}={\bf
N}|_{{\bf R}\equiv0}, \ S=B|_{{\bf R}\equiv0}.$

The vectors $\tilde {\bf E},\ \tilde {\bf H}$ and scalar $S$ belong
to a carrier space of the representation $D_1(2,1,1)$. Their Galilei
transformation laws look as \beq\label{9}\bea{l}\tilde {\bf
E}\to\tilde {\bf E}+{\bf v}\times \tilde {\bf H}+{\bf v}S,\\\tilde
{\bf H}\to\tilde {\bf H},\ S\to S.\eea\eeq

In accordance with (\ref{9}) $S$ belongs to invariant subspace of
Galilean transformations, so we can impose one more Galilei
invariant condition \beq\label{10}S=0\ \ \texttt{or }\frac{\partial
A^4}{\partial t}=0.\eeq As a result we come to equations
(\ref{mag}), i.e., to the magnetic limit of Maxwell's equations.
Thus equations (\ref{mag}) are nothing but the system of equations (\ref{coupl})
with the additional Galilei invariant constrains (\ref{V=0}) and (\ref{10}).

Galilei transformations for solutions of equations (\ref{mag}) are
given by equations (\ref{tran}). Again we recognize the invariant
subspace spanned on vectors ${\bf H}_m$, and so it is possible to
impose the invariant condition \beq \label{17}{\bf H}_m=0 \ \ \texttt{or
}{\bf A}=\nabla \varphi,\ \ {\bf j}_m=0\eeq where $\varphi$ is a
solution of the Laplace equation. As a result we come to the
following system \beq\label{mag1}\bea{l}\nabla\times \widehat{\bf
E}=0, \ \ \nabla\cdot\widehat{\bf E}=e\rho\eea\eeq where we denote
$\widehat{\bf E}={\bf E}_m|_{{\bf H}_m\equiv0}$ and
$\rho=j^0_m|_{{\bf H}_m\equiv0}.$

Equation (\ref{mag1}) is still Galilei invariant, moreover, both
$\widehat{\bf E}$ and $\rho$ are not changing under the Galilei
transformations. The related potential $\tilde A$ is constrained by
conditions (\ref{V=0}),(\ref{10}) and
(\ref{17}). Moreover, up to gauge transformations it is possible to set
in  (\ref{17}) ${\bf A}=0$.

\subsection{Other reductions}

Equations (\ref{coupl1}), (\ref{mag}) and (\ref{mag1}) exhaust all Galilei
invariant systems which can be obtained starting with (\ref{coupl})
and imposing additional constraints which reduce the number of dependent
variables. To find other Galilei invariant equations for massless vector
fields we use the observation
 that the Galilei transformations for equations (\ref{coupl}) have the form
 given by relations  (\ref{fin}) and (\ref{16}) if we change
 ${\bf N}\to {\mbox{\boldmath$\cal N$\unboldmath}},
 {\bf W}\to{\mbox{\boldmath$\cal W$\unboldmath}}, \cdots $ here.
 Thus the subsystem of equations (\ref{coupl}) obtained by excluding
 equations ${\mbox{\boldmath$\cal N$\unboldmath}}=0$ and ${\cal J}^0=0$:
\beq\label{coupl2}\bea{l}{\mbox{\boldmath${\cal J}$\unboldmath}}
\equiv \nabla\times{\bf W}+\frac{\partial}{\partial t}{\bf R}-e{\bf
j}=0,\\{\cal J}^4\equiv\nabla\cdot{\bf
R}-ej^4=0,\\{\mbox{\boldmath${\cal
W}$\unboldmath}}\equiv\frac{\partial}{\partial t}{\bf R}-\nabla
B=0,\\ {\mbox{\boldmath${\cal R}$\unboldmath}}\equiv \nabla\times
{\bf R}=0,\\ {\cal B}\equiv\nabla\cdot {\bf W}=0\eea\eeq is Galilei
covariant too and does not include dependent variables $\bf N$ and
$j^0$. The Galilean transformations for ${\bf W}, {\bf R}, B$ and
${\bf j}, j^4$ are still given by equations (\ref{fin}) and
(\ref{16}).

Following the analogous reasonings it is possible to exclude from
(\ref{coupl2}) equations ${\mbox{\boldmath$\cal J$\unboldmath}}=0$ and
${\cal B}=0$ and obtain the system
 \beq\label{coupl4}\bea{l}\nabla\cdot{\bf
R}-ej^4=0,\\\frac{\partial} {\partial t}{\bf R}-\nabla B=0,\\
\nabla\times {\bf R}=0\eea\eeq which includes only two vector and
two scalar variables. The related potential without loss of
generality reduces to the only variable $A^4$.

The other way to reduce system (\ref{coupl2}) is to exclude the
equation ${\mbox{\boldmath${\cal W}$\unboldmath}}=0$.  As a result
we come to the electric limit for the Maxwell equation (\ref{el})
for ${\bf W}={\bf H}_e$ and ${\bf R}={\bf E}_e$.

Thus in addition to (\ref{coupl}), (\ref{coupl1}), (\ref{mag}) and
(\ref{mag1}) we have three more Galilei invariant systems given by
equations (\ref{coupl2}), (\ref{coupl4}) and (\ref{el}). These
equations admit additional reductions by imposing Galilei invariant
constrains to their solutions.

Considering (\ref{el}) we easily find a possible invariant
conditions ${\bf E}_e=0, \ j^4_e=0$ which reduce it to the following
equations \beq\label{111}\nabla\times\hat{\bf H}=e{\bf j},\ \
\nabla\cdot\hat{\bf H}=0\eeq where we denote $\hat{\bf H}={\bf
H}_e|_{{\bf E}_e\equiv0}$.

Let us return to system (\ref{coupl2}). This system can be reduced
by imposing the Galilei invariant conditions ${\bf R}=0, j^4=0$ to
the following form: \beq\label{coupl3}\bea{l}\nabla\times\hat{\bf
H}-e{\bf j}=0,\\\nabla\cdot \hat{\bf H}=0,\ \nabla S=0 \eea\eeq
where we change the notations ${\bf W}\to\hat{\bf H}$ and $B\to S$. The related
potential have the form $A=(A^4,0,{\bf A})$, moreover, $A^4$ should
satisfy the condition $\nabla A^4=0$.

We see that in contrast with the relativistic theory there exist a
big variety of equations for massless vector fields invariant with
respect to the Galilei group. The list of such equations is given by
formulae (\ref{mag}), (\ref{el}), (\ref{coupl}), (\ref{coupl1}),
(\ref{mag1})--(\ref{coupl3}).

\section{Discussion}

The revision of classical results \cite{lebellac} related to
Galilean electromagnetism made in the present paper appears to be
successful. Our knowledge of indecomposable representations of
homogeneous Galilei group defined in vector and scalar fields
\cite{NN1} made it possible to complete the results of Le Bellac and
L\'evy-Leblond \cite{lebellac} and present an extended class of
Galilei-invariant equations for massless vector fields. Among them
are decoupled systems of first order equations which include the
same number of components as the Maxwell equations and equations
with other numbers of components as well. The most extended system
includes ten components while the most reduced one has three
components only.

It is necessary to stress that the majority of obtained equations
admit clear physical interpretations. Thus equation (\ref{mag1}) and
(\ref{111}) are applied in electro- and magnetostatics
respectively.
   The very procedure of
deducing of Galilei invariant equations for vector fields used in
the present paper makes their interpretation to be rather
straitforward, since any obtained equation has its relativistic
counterpart.

We see that the number of Galilean wave equations for massless
vector fields is rather extended, and so there are many
possibilities to describe interaction of non-relativistic charged
particles with external gauge fields. Some of these possibilities
have been discussed in \cite{NN1} and \cite{NN3}, se also
\cite{FN2}, \cite{NF3} and \cite{ijtp}, \cite{santos}. Starting with
the found equations and using the list of functional invariants for
Galilean vector fields presented in \cite{NN2} it is easy to
construct nonlinear models invariant with respect to the Galilei
group, including its supersymmetric extensions.

Invariance of any particular found equation with respect to the
Galilei transformations can be verified by direct calculation. The
main result presented in this paper is the completed description of
all such equations.


\begin{thebibliography}{99}

\bibitem{lebellac}
Le Bellac M and L\'evy-Leblond J M \NCB{Galilean
electromagnetism}{1973}{14}{217-33}

\bibitem{Hol} Holland P and Brown H R 2003 The Non-Relativistic Limits
of the Maxwell and Dirac Equations: The Role of Galilean and Gauge
Invariance {\it Studies in History and Philosophy of Science}  {\bf
34} 161-87

\bibitem{NN1} de Montigny M,  Niederle J and Nikitin A G \JPA
{Galilei invariant theories. I. Constructions of indecomposable
finite-dimensional representations of the homogeneous Galilei group:
directly and via contractions}{2006}{39}{1-21}


\bibitem{ijtp} de Montigny M, Khanna F C and Santana A E
\IJTP{Nonrelativistic wave equations with gauge
fields}{2003}{42}{649-71}

\bibitem{santos} Santos E S, de Montigny
M, Khanna F C and Santana A E \JPA{Galilean covariant Lagrangian
models}{2004}{37}{9771-91}





\bibitem{FN1} Fushchich V I and Nikitin A G
\JPA{Reduction of the representations of the generalised Poincar\'e
algebra by the Galilei algebra}{1980}{13}{2319-30}

\bibitem{FN2} Fushchich W I and Nikitin A G 1994 {\em
Symmetries of Equations of Quantum Mechanics} (New York: Allerton
Press)

\bibitem{contraction} In\"on\"u E and Wigner E P
\PNASUS{On the contraction of groups and their
representations}{1953}{39}{510-24}

\bibitem{NN2} Niederle J and  Nikitin AG 2006 Construction and
classification of indecomposable finite-dimensional representations
of the homogeneous Galilei group {\it Czechoslovak
     Journal of Physics} {\bf 56} 1243-50


\bibitem{NN3} Niederle J and Nikitin A G
{Galilei invariant theories. II.
Wave equations for massive fields} math-phys???


\bibitem{ll1967} L\'evy-Leblond J M \CMP{Non-relativistic particles
and wave equations}{1967}{6}{286-311}

\bibitem{levyleblond} L\'evy-Leblond J M 1971 Galilei group and
galilean invariance, in {\it Group Theory and Applications} Ed. E.M.
Loebl, Vol. II (New York: Academic) 221-99

\bibitem{levycarroll}
Wightman A S 1959 Relativistic Invariance and Quantum Mechanics (Notes by A. Barut) {\it Nuovo Cimento Suppl}, {\bf 14}, 81-94;
\\Hamermesh M \AP{Galilean invariance and the Schr\"odinger equation}{1960}
{9} {518-521}

\bibitem{NF3} Nikitin A G and
Fuschich W I \TMP{Equations of motion for particles of arbitrary
spin invariant under the Galilei group}{1980}{44}{584-92}





















\end{thebibliography}
\end{document}